# Laser-Synthesized Ligand-Free Cu Nanocatalysts in Electrochemical $CO_2$ Reduction to Methane

*Taiping Ye[1], Artyom Lactionov[2], Islam Sozaev[2], Anton Popov[2], Sergei Klimentov[2], Andrei V. Kabashin [3\*], and Ya Liu[1\*]*

1 - International Research Center for Renewable Energy, State Key Laboratory of Multiphase Flow in Power Engineering, Xi'an Jiaotong University, Shaanxi 710049, China

2 - Laboratory 'Bionanophotonics', Institute of Engineering Physics for Biomedicine (PhysBio), National Research Nuclear University MEPhI, Moscow 115409, Russia

3 - CNRS, LP3, Aix-Marseille Université, 13288 Marseille, France

\* Corresponding authors:
Ya Liu : yaliu0112@xjtu.edu.cn
Andrei V. Kabashin : kabashin@lp3.univ-mrs.fr

ABSTRACT:

Electrochemical $CO_2$ reduction (e$CO_2$R) represents a pivotal strategy for mitigating global carbon emissions while simultaneously converting renewable energy into storable chemical fuels. Copper-based catalysts have been extensively explored in this field due to their unique capability to catalyze multi-carbon products. However, the intrinsic complexity of e$CO_2$R pathways on Cu surfaces often leads to mixed product distributions, posing a significant challenge for achieving high selectivity toward a single desired hydrocarbon. Herein, we report a breakthrough in methane selectivity using laser-synthesized, ligand-free Cu nanomaterials. Unlike conventional Cu catalysts that produce diverse products, these ligand-free nanoparticles exhibit unprecedented selectivity for methane ($CH_4$) with a Faradaic efficiency (FE) exceeding 70% at superior overpotentials. The absence of surface ligands, a direct consequence of the ultrafast laser ablation synthesis, ensures abundant exposed active sites with tailored electronic and geometric configurations. We attribute the exceptional methane selectivity to the synergistic effects of active sites-rich surfaces and optimized *CO intermediate binding energetics, which favor the protonation pathway toward $CH_4$ rather than C–C coupling. This work not only resolves the long-standing selectivity dilemma in Cu-catalyzed e$CO_2$R but also establishes laser-synthesized ligand-free nanomaterials as a versatile platform for designing high-performance electrocatalysts.

**KEYWORDS:** e$CO_2$R, laser ablation in liquids, nanoparticles, copper, methane selectivity

## INTRODUCTION

The escalating atmospheric $CO_2$ concentration, driven by fossil fuel combustion, has intensified global warming and environmental degradation.[1-3] Electrochemical $CO_2$ reduction (e$CO_2$R) offers a dual-purpose solution: curtailing greenhouse gas levels and storing intermittent renewable energy in the form of chemical bonds within portable fuels.[4-8] Among potential products, methane ($CH_4$) stands out as an energy-dense fuel compatible with existing gas infrastructure.[9-11] Methane exhibits seamless compatibility with existing natural gas infrastructure, enabling a smooth transition toward adopting e$CO_2$R-derived methane.[12, 13] Consequently, the development of advanced electrocatalysts for achieving high-selectivity $CO_2$-to-$CH_4$ conversion represents a critical research imperative.[14-16]

Copper has maintained its predominant position in electrocatalytic $CO_2$ reduction research owing to its distinctive capability to direct multi-carbon product synthesis through C–C coupling pathways.[17-20] Foundational studies have unveiled the inherent catalytic complexity of Cu-based systems: Polycrystalline Cu surfaces simultaneously generate C1 products (CO, $CH_4$) and C2+ compounds, with the selectivity of these dual reaction pathways exhibiting marked dependence on the surface lattice orientation, dynamic redox states, and interfacial microenvironment.[21-24] Although nanostructured or alloy systems have achieved moderate selectivity efficiencies, critical performance barriers remain unbroken. Notably, methane—the thermodynamically favored product—persistently struggles to surpass the 50% Faradaic efficiency threshold in Cu-based systems.[18, 25] A more fundamental challenge arises from the dynamic reconstruction behavior of Cu catalytic interfaces under operational potentials, which continuously modifies the electronic configuration of active sites.[26-28] This inherent instability creates an intrinsic conflict between the imperative to precisely regulate critical intermediate adsorption strengths and the need to suppress C–C coupling events.

Therefore, the crucial strategy is to solve mutually restrictive objectives that fundamentally constrain the directional enhancement of methane selectivity. Laser ablation in liquids (LAL) emerged as a facile synthesis method, enabling one-step fabrication of ligand-free, ultrapure nanoparticles (NPs).[29-31] When employed in the ultrashort (femtosecond) laser ablation regime, the LAL technique requires significantly lower energy per pulse compared to conventional long-pulse ablation, which can drastically reduce detrimental plasma and cavitation effects broadening size distribution.[32] This reduction in energy input enables finer control over nanoparticle size and morphology while also allowing synthesis in a broad range of organic solvents without their decomposition. As a result, the physico-chemical characteristics of the produced NPs can be precisely tuned by adjusting the ablation parameters and solvent environment,[33-35] which renders possible the growth of nanomaterials with engineered properties, including the production of metastable phases and defect-rich surfaces.[36, 37] This property provides the possibility to fundamentally address the limitation of methane directional selectivity. The non-equilibrium nature of laser ablation preserves crystallographic irregularities—such as steps, kinks, and vacancies—that could serve as catalytic hotspots.[38, 39]

Herein, we leverage LAL to synthesize ligand-free Cu nanoparticles (LF-Cu NPs) with engineered surface defects for selective $CO_2$-to-$CH_4$ conversion. Unlike traditional Cu catalysts, LF-Cu NPs exhibit a record-breaking $CH_4$ FE of 70%, simultaneously maintaining >90% initial activity after 30 h. By correlating synthesis parameters (laser fluence, pulse duration) with catalytic performance, we establish design principles for LF-Cu NPs $eCO_2R$ catalysts. This study transcends conventional catalyst optimization by integrating laser nanotechnology with $CO_2$ electrochemistry.

The elimination of ligands not only maximizes active site utilization but also circumvents stability issues plaguing colloidal nanocatalysts. Our findings underscore the untapped potential of non-equilibrium synthesis routes in unlocking unprecedented selectivity regimes, offering a blueprint for next-generation catalysts tailored for complex $eCO_2R$. As renewable energy capacities expand globally, such scalable and selective systems will prove indispensable for closing the carbon cycle.

## EXPERIMENTAL

**Fabrication of LF-Cu NPs.** The NPs were synthesized using femtosecond LAL, following a procedure similar to our previous studies. We employed a ytterbium laser with the following parameters: wavelength 1030 nm, pulse duration 270 fs, pulse energy 30 μJ, and repetition rate 200 kHz. The laser radiation was focused on a metallic copper target using a 100 mm working distance F-theta objective. The target was mounted vertically in an ablation chamber filled with isopropanol. To optimize the synthesis productivity, the liquid thickness along the laser pathway was maintained at 3 mm. Following the synthesis, the NPs were concentrated to a concentration of 1 mg/ml by centrifugation.

**Characterization of LF-Cu NPs.** The size and surface morphology of the LF-Cu NPs were analyzed using scanning electron microscopy (SEM) with an MAIA 3 microscope (Tescan) operating at 30 kV accelerating voltage. Energy-dispersive spectrometry (EDS) using an X-act detector (Oxford Instruments) was conducted to study the qualitative chemical composition of the NPs. For the SEM and EDS analysis, samples were prepared by depositing 5 μl of the nanoparticle colloid onto a cleaned monocrystalline silicon wafer, followed by drying under ambient conditions. Size distributions were determined by measuring the sizes of more than 500 individual NPs from SEM images.

Hydrodynamic sizes and zeta-potential of the NPs were measured using dynamic light scattering in the Smoluchowski approximation with a Zetasizer Nano-ZS (Malvern Instruments). Optical extinction spectra of the synthesized nanoparticles were recorded in glass cuvettes with a 10 mm optical path using an MC-111 spectrophotometer (Sol instruments).

**Fabrication of GDEs Samples.** The LF-Cu NPs were dispersed in the isopropanol solvent and sonicated for 30 minutes to form a homogeneous ink. Subsequently, the prepared ink was then spray-coated onto the commercial carbon paper using an airbrush and dried at 70°C for 1 h to ensure complete solvent removal. In addition, the catalyst loading was controlled at 0.5 mg cm$^{-2}$ to ensure uniformity and reproducibility.

**Electrochemical $CO_2$ reduction measurements.** The $CO_2$ reduction reactions were conducted with a self-built electrolysis system (Figure 1). Based on the three-electrode system, an electrochemical workstation is used to conduct electrochemical tests on the flow cell. The flow cell is composed of the catholyte chamber, the anolyte chamber, the gas flow plate, electrodes, ion exchange membranes, and sealing accessories. The prepared GDE, leak-free Ag/AgCl, and platinum net were used as the working electrode, reference electrode, and counter electrode, respectively.

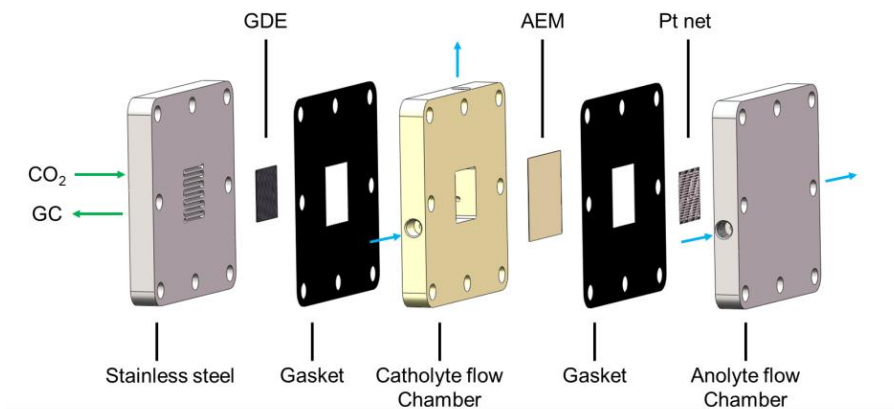

Figure 1. Schematic diagram of three-electrode flow cell configuration

## RESULTS AND DISCUSSION

**Structure of LF-Cu NPs.**

The LF-Cu NPs were synthesized in isopropanol. This choice of solvent is significant because it helps to minimize the oxidation of the NPs surface during LAL synthesis compared to ablation in water. Colloidal solutions of LF-Cu NPs prepared in isopropanol exhibited a dark-brown color, whereas those prepared in water (data not shown) had a green tint, which is characteristic of oxidized copper. This observation indirectly confirmed that the LF-Cu NPs synthesized in isopropanol were less oxidized. To further investigate the chemical composition of the NPs synthesized in isopropanol, we performed an EDS analysis (Figure 2c). The results revealed strong peaks attributed to copper and silicon. The signal from the silicon is due to the silicon substrate used for sample preparation, while the strong copper signal could only be attributed to the LF-Cu NPs. We also identified carbon and a very weak oxygen signal. The presence of carbon is typical for all EDS measurements and cannot be directly attributed to the carbonization of the NPs, although this possibility cannot be completely excluded. The very low oxygen signal confirms that the NPs were almost unoxidized. In summary, EDS analysis indicates that the LF-Cu NPs prepared in isopropanol consist of almost pure elemental copper with minor oxidation. While the possibility of partial carbonisation cannot be completely ruled out, the overall composition suggests high purity. For the eCO$_2$R tests, all NPs were prepared in isopropanol, and the subsequent results are provided for the LF-Cu NPs prepared in this solvent.

Electron microscopy (Figure 2a) revealed that the synthesized LF-Cu NPs have a spherical morphology and a lognormal size distribution with a mode size of 23 nm. The spherical shape is typical for NPs prepared by LAL synthesis. The hydrodynamic size distribution (Figure 2b) of the LF-Cu NPs showed a larger mode size of 37 nm. This moderate difference between the physical (measured by electron microscopy) and hydrodynamic sizes can be attributed to the formation of a layer of adsorbed molecules and associated liquid molecules, which contribute to the measured hydrodynamic size. Overall, the good agreement between the physical and hydrodynamic sizes confirms that the LF-Cu NPs are colloidally stable and do not form aggregates in liquids. This high colloidal stability of the laser-synthesized LF-Cu NPs is generally attributed to the high electrostatic charging of the ablated NPs.

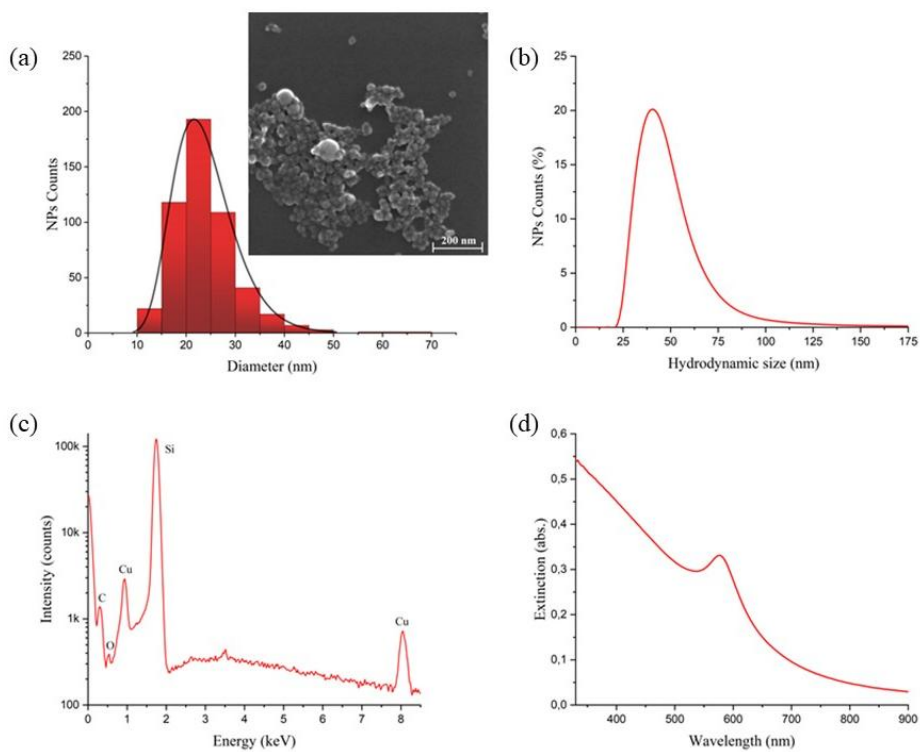

Figure 2. Physicochemical characterization of LF-Cu NPs. (a) Size distribution with a typical SEM image. (b) Distribution of hydrodynamic sizes. (c) EDS specter, demonstrating qualitative chemical composition. (d) Optical extinction spectrum.

The optical extinction spectrum (Figure 2c) exhibited a plasmonic peak centered around 575 nm. The presence of this plasmonic peak serves as further confirmation of the metallic nature of the synthesized LF-Cu NPs and indicates that their surface has low oxidation.

**$CO_2$ reduction performance of LF-Cu NPs**

Prior to the formal evaluation of the synthesized LF-Cu NPs, we systematically assessed the $CO_2$ reduction potential of commercial copper NPs (25 nm). Figure 3 presents the $CO_2$ reduction performance of the commercial copper catalyst in 1M KOH electrolyte. The results reveal complex product distribution with unsatisfactory selectivity toward individual products. The system exhibited an overall transition trend from C1 products to multi-carbon (C2+) products with increasing current density. Notably, a peak Faradaic efficiency of 43.02% for CO was observed at 100 mA cm$^{-2}$ current density, while $C_2H_4$ production demonstrated a maximum Faradaic efficiency of 39.01% at 400 mA cm$^{-2}$.

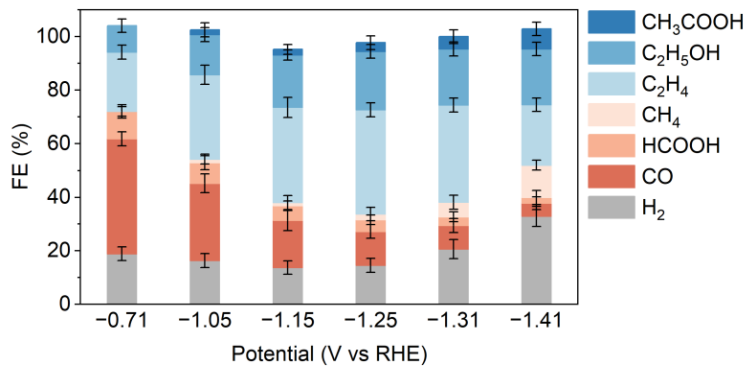

Figure 3. eCO$_2$R performance evaluation of commercial Cu catalysts in flow cell

Further building on these findings, the LF-Cu NPs synthesized by LAL were systematically evaluated, with corresponding results presented in Figure 4. Notably, the LF-Cu NPs exhibited fundamentally distinct catalytic behavior compared to commercial copper counterparts. The electrochemical analysis revealed exceptional selectivity toward $CH_4$ production, with the Faradaic efficiency reaching a remarkable peak value of 67.01% at an applied current density of 600 mA cm$^{-2}$. This performance benchmark establishes new international leadership in $CH_4$ among copper-based catalytic systems for $CO_2$ reduction applications.

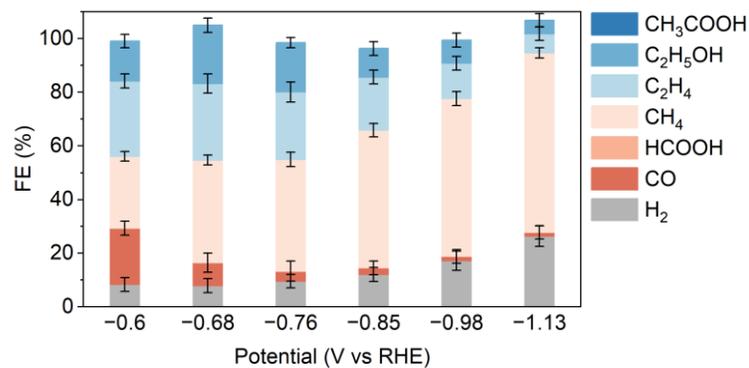

Figure 4. eCO$_2$R performance evaluation of LF-Cu NPs catalysts in flow cell

To elucidate the underlying mechanism responsible for this enhanced performance, we conducted electrochemical cyclic voltammetry (CV) measurements of two different copper catalysts (shown in Figure 5). The experimental data revealed that the LF-Cu NPs exhibited substantially enhanced current response characteristics compared to conventional commercial copper catalysts, demonstrating a 4.2-fold increase in cathodic current density at -1.6 V vs Ag/AgCl.

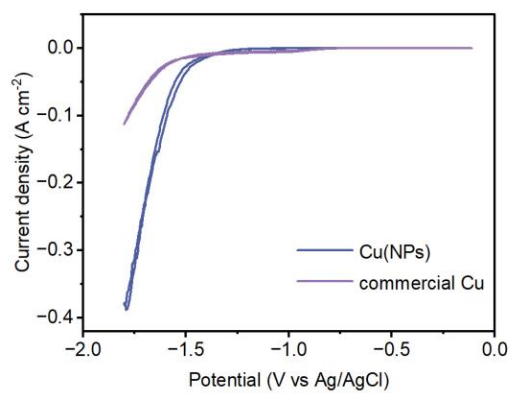

Figure 5. Electrochemical CV characterization test results

# CONCLUSIONS

The development of laser-synthesized ligand-free copper nanomaterials (LF-Cu NPs) represents a transformative advancement in electrocatalytic $CO_2$ reduction, specifically addressing the long-standing selectivity challenges of Cu-based catalysts. By circumventing the limitations of conventional synthesis methods, which rely on organic ligands or post-synthesis treatments, this work pioneers a one-step laser ablation strategy to fabricate defect-rich, ligand-free Cu NPs with unpassivated active sites. The elimination of surface ligands not only enhances charge transfer kinetics but also prevents active site poisoning—a critical factor often overlooked in colloidal nanocatalyst design.

The intrinsic structural features of LF-Cu NPs, including coordinatively unsaturated Cu atoms and stabilized grain boundaries, are shown to play a decisive role in redirecting the $CO_2$ reduction pathway toward methane. Unlike traditional Cu catalysts, where dynamic surface reconstruction under operational conditions leads to unpredictable product distributions, the laser-synthesized LF-Cu NPs retain a static, defect-dominated architecture. This structural permanence enables sustained methane selectivity (>70% Faradaic efficiency) and operational stability over 30 hours, outperforming state-of-the-art Cu-based systems. Mechanistically, in-situ spectral characterization reveals that the synergy between ligand-free surfaces and tailored defect geometries lowers the kinetic barrier for stabilizing the *CHO intermediate—a pivotal methane precursor—while disfavoring C–C coupling pathways. These insights redefine the design principles for $eCO_2R$ catalysts, shifting the focus from sole optimization of *CO binding energy to a dual emphasis on defect engineering and surface cleanliness.

This study underscores the broader potential of laser-synthesized, ligand-free nanomaterials in electrocatalysis. The scalability of the synthesis method, coupled with its compatibility with diverse metallic Cu targets, offers a versatile platform for designing catalysts for complex multi-step reactions. By bridging non-equilibrium nanomaterial synthesis with atomic-level mechanistic understanding, this work provides a blueprint for advancing $CO_2$ valorization technologies toward industrial viability. Future efforts should explore the extension of this paradigm to multi-metallic systems and its integration with renewable energy infrastructures, ultimately accelerating the transition to a closed-loop carbon economy.

## ASSOCIATED CONTENT

**Supporting Information.** Detailed fabrication procedures; supplemental structural characterization data; supplemental photoelectrochemical measurement data; efficiency evaluation methods.

## AUTHOR INFORMATION

**Corresponding Authors**

*yaliu0112@xjtu.edu.cn,

*kabashin@lp3.univ-mrs.fr

**Author Contributions**

The manuscript was written through contributions of all authors. All authors have given approval to the final version of the manuscript.

## ACKNOWLEDGMENT

This work is supported by the National Key R&D Program of China (2024YFF0506100), the Sichuan Science and Technology Program (No. 2024YFHZ0037), the Natural Science Basic Research Program of Shaanxi (No. 2024JC-YBMS-284), the Key Research and Development